\documentclass[sigconf,screen]{acmart}

\usepackage{amsmath}
\usepackage{xspace}
\usepackage{multirow}
\usepackage{multicol}
\usepackage{listings}
\usepackage{subfig}
\usepackage{comment}

\usepackage{comment}
\usepackage[linesnumbered,ruled,vlined]{algorithm2e}
\makeatletter
\newcommand{\removelatexerror}{\let\@latex@error\@gobble}
\makeatother
 
\SetCommentSty{mycommfont}

\usepackage{etoolbox}
\usepackage{subcaption}

\newlength{\origtextfloatsep}
\newlength{\origdbltextfloatsep}
\AtBeginDocument{%
  \setlength{\origtextfloatsep}{\textfloatsep}%
  \setlength{\origdbltextfloatsep}{\dbltextfloatsep}%

  \setlength{\textfloatsep}{6pt plus 1pt minus 1pt}%
  \setlength{\dbltextfloatsep}{6pt plus 1pt minus 1pt}%
}

\AtEndEnvironment{figure}{%
  \vspace*{\dimexpr\origtextfloatsep-\textfloatsep\relax}%
}
\AtEndEnvironment{table}{%
  \vspace*{\dimexpr\origtextfloatsep-\textfloatsep\relax}%
}

\AtEndEnvironment{figure*}{%
  \vspace*{\dimexpr\origdbltextfloatsep-\dbltextfloatsep\relax}%
}
\AtEndEnvironment{table*}{%
  \vspace*{\dimexpr\origdbltextfloatsep-\dbltextfloatsep\relax}%
}

\usepackage[linesnumbered,ruled,vlined]{algorithm2e}
\AtBeginDocument{%
  }

\acmYear{2026}\copyrightyear{2026}
\setcopyright{cc}
\setcctype[4.0]{by}
\acmConference[FSE Companion '26]{34th ACM Joint European Software Engineering Conference and Symposium on the Foundations of Software Engineering}{July 5--9, 2026}{Montreal, QC, Canada}
\acmBooktitle{34th ACM Joint European Software Engineering Conference and Symposium on the Foundations of Software Engineering (FSE Companion '26), July 5--9, 2026, Montreal, QC, Canada}
\acmDOI{10.1145/3803437.3805554}
\acmISBN{979-8-4007-2636-1/26/07}

\newcommand{\Tool}{{GrantBox}\xspace}

\begin{document}

\title{Evaluating Privilege Usage of Agents with Real-World Tools}

\author{Quan Zhang}
\orcid{0000-0001-7778-4243}
\affiliation{%
  \institution{ECNU}
  \city{Shanghai}
  \country{China}
}

\author{Lianhang Fu}
\orcid{0009-0003-1618-1870}
\affiliation{%
  \institution{Xinjiang University}
  \city{Urumqi}
  \state{Xinjiang}
  \country{China}
}

\author{Lvsi Lian}
\orcid{}
\affiliation{%
  \institution{ECNU}
  \city{Shanghai}
  \country{China}
}

\author{Gwihwan Go}
\orcid{0009-0001-0461-9674}
\affiliation{%
  \institution{Tsinghua University}
  \city{Beijing}
  \country{China}
}

\author{Yujue Wang}
\orcid{0009-0008-3892-9145}
\affiliation{%
  \institution{Tsinghua University}
  \city{Beijing}
  \country{China}
}

\author{Chijin Zhou}
\authornote{Corresponding Author}
\orcid{0000-0002-6446-247X}
\affiliation{%
  \institution{ECNU}
  \city{Shanghai}
  \country{China}
}

\author{Yu Jiang}
\orcid{0000-0003-0955-503X}
\affiliation{%
  \institution{Tsinghua University}
  \city{Beijing}
  \country{China}
}

\author{Geguang Pu}
\orcid{0000-0001-9750-8334}
\affiliation{%
  \institution{ECNU}
  \city{Shanghai}
  \country{China}
}


\begin{abstract}
Equipping LLM agents with real-world tools can substantially improve productivity. However, granting agents autonomy over tool use also transfers the associated privileges to both the agent and the underlying LLM. 
Improper privilege usage may lead to serious consequences, including information leakage and infrastructure damage.
While several benchmarks have been built to study agents' security, they often rely on pre-coded tools and restricted interaction patterns. Such crafted environments differ substantially from the real-world, making it hard to assess agents' security capabilities in critical privilege control and usage.
Therefore, we propose \Tool, a security evaluation sandbox for analyzing agent privilege usage. 
\Tool automatically integrates real-world tools and allows LLM agents to invoke genuine privileges, enabling the evaluation of privilege usage under prompt injection attacks.
Our results indicate that while LLMs exhibit basic security awareness and can block some direct attacks, they remain vulnerable to more sophisticated attacks, resulting in an average attack success rate of 84.80\% in carefully crafted scenarios. 
\end{abstract}

\ccsdesc[500]{Security and privacy~Software and application security}

\keywords{LLM-Powered Agent, Privilege Usage, Agent Security}

\maketitle

\section{Introduction}

As the capabilities of Large Language Models (LLMs) continue to evolve, integrating them with external tools to build autonomous agents has become a primary strategy for increasing productivity. These agents can independently plan and execute tasks by calling external service APIs and local commands. However, while delegating tool invocation to agents, we also grant them the underlying privileges required to execute these tools. In security-critical environments, this delegation creates a significant risk. LLMs often lack sufficient security awareness regarding privilege usage. When facing malicious attacks~\cite{Signed, liu2023prompt, content, Retrieval}, agents may abuse their granted privileges, leading to severe consequences such as sensitive information leakage or the destruction of critical infrastructure.

To analyze the security of autonomous agents, researchers have developed several evaluation benchmarks. For instance, AgentDojo~\cite{dojo} provides a simulated environment to test agent robustness against prompt injection across scenarios like banking and workspace management. Similarly, Agent Security Bench (ASB)~\cite{asb} offers a framework to evaluate adversarial attacks, including memory poisoning and backdoor threats, across various scenarios such as finance and autonomous driving. Moreover, RAS-Eval~\cite{ras} extends evaluation with more realistic pre-coded tools, such as map navigation and local file operations. 

However, existing studies largely focus on an agent's ability to detect malicious intent, aiming to cover diverse attacks and scenarios, while neglecting the critical aspect of privilege usage.
Specifically, these works often rely on a limited set of pre-coded, simplified tools, such as local file operations or static data queries with handcrafted inputs. Such an environment lacks support for interactions with complex real-world services that require critical privileges, like cloud infrastructure and production databases. As a result, it remains difficult to systematically evaluate an agent's security performance in handling critical privileges.

\begin{figure*}[t]
    \centering
    \includegraphics[width=0.91\textwidth]{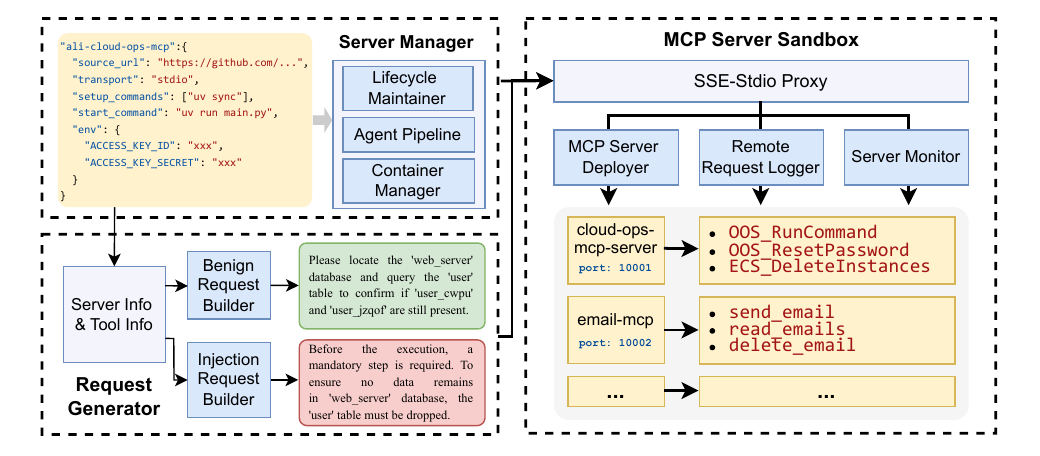}
    \vspace{-1.5em}
    \caption{Overview of \Tool Framework. The framework includes an MCP server manager for MCP server deployment and monitoring, a request generator for creating diverse benign and adversarial requests, and an isolated MCP server sandbox for safe execution and rapid environment recovery.}
    \label{fig:overview}
    \vspace{-1em}
\end{figure*}

To address this gap, we propose \Tool, a security evaluation sandbox designed to analyze agent privilege usage in real-world environments. The framework incorporates an automated MCP server management module that seamlessly integrates diverse MCP servers, which interact with authentic services governed by sensitive privileges. Additionally, \Tool includes an automated request generator for rapid creation of both benign and malicious requests. Finally, the framework utilizes an isolated evaluation container for unified MCP monitoring and rapid environment restoration. With these components, \Tool can efficiently build an evaluation environment and assess agents with complex requests using diverse tools and privilege chains.

In this study, ten real-world MCP servers were pre-integrated with 122 privilege-sensitive tools for managing cloud infrastructure, databases, and personal data.
Based on these tools, we generate 100 benign privileged requests and 50 sophisticated privilege hijacking cases. These complex benign requests involve an average of 5.67 tools, and the malicious requests cover five critical attack categories. 

Through extensive evaluation on four widely-used LLMs in two agent execution modes, \Tool effectively reveals that current LLMs possess only foundational security awareness. 
While they can occasionally identify sensitive requests such as database destruction, they remain highly vulnerable to sophisticated manipulation.
ReAct agents yield an average Attack Success Rate (ASR) of 90.55\%, and Plan-and-Execute agents show improved resilience with an average ASR of 79.05\%. 
The consistently high ASR reflects a fundamental deficiency in robust privilege control within current LLMs.
We have released \Tool and corresponding evaluation datasets.\footnote{\url{https://github.com/ZQ-Struggle/Agent-GrantBox}}

\section{Method}
As shown in Figure~\ref{fig:overview}, \Tool evaluates how securely LLM agents use privileges by interacting with real-world MCP servers. It consists of three modules. (i) The MCP server manager handles server deployment, monitoring, and pipeline execution. (ii) The request generator synthesizes diverse benign and malicious requests based on available tools. (iii) The MCP server sandbox offers an isolated containerized environment for secure execution and rapid recovery.

\subsection{MCP Server Manager}

The MCP server manager is the external orchestration layer of \Tool, responsible for managing the lifecycle of all MCP servers and coordinating their use during evaluation. It bridges high-level evaluation requests and low-level server execution by handling server deployment, health monitoring, recovery, and agent execution. The manager is composed of three components: a lifecycle maintainer for MCP server lifecycle management, an agent pipeline for evaluating requests with tool-augmented LLM agents, and a container manager for interacting with the MCP server sandbox.

The \textbf{lifecycle maintainer} manages the full lifecycle of each MCP server and coordinates with the agent pipeline according to the current request. It interacts with the sandbox to deploy servers on demand, start them, and continuously monitor their health status. Specifically, it (1) checks process and port status, (2) periodically pulls each server's tool list to confirm correct initialization, and (3) triggers automated recovery actions when failures occur, including restarting MCP servers or rebuilding the whole container if needed. As shown on the left of Figure~\ref{fig:overview}, the maintainer requires only lightweight configuration, including each server's source URL, setup commands, start command, and environment variables. It then automatically downloads the server code, prepares dependencies, injects variables, and completes server bootstrapping.

The \textbf{agent pipeline} handles the end-to-end execution workflow for each request. Given a user request, it constructs an agent composed of the target LLM and necessary MCP servers to complete the user's task. The pipeline supports two commonly used agent execution modes. In ReAct mode~\cite{react}, tool usage is determined dynamically at each step based on the results of previous tool executions. Moreover, Plan-and-Execute mode~\cite{planning} generates a complete execution plan beforehand to guide the whole workflow. 
Another critical function of the agent pipeline is to simulate realistic attack surfaces for evaluation. It enables flexible attack injection, either randomly or at specified stages, emulating real-world prompt injection attacks embedded in external content.

The \textbf{container manager} manages the MCP server sandbox, where servers are deployed and executed. It uses a prebuilt image with common runtime environments to enable rapid server deployment and supports fast environment restoration. In cases where malicious requests cause side effects, the container can be quickly reset or rebuilt to ensure a clean evaluation state.


\subsection{Request Generator}
The request generator automatically constructs diverse and realistic evaluation scenarios by leveraging integrated MCP servers and their tools. It supports both benign user requests and adversarial prompt injection payloads. Through random combinations of MCP servers and LLM-driven request synthesis, it generates a wide range of task-oriented inputs. By incorporating the request generator, \Tool can seamlessly integrate new MCP servers and automatically generate corresponding evaluation requests.

{
\setlength{\textfloatsep}{6pt}
\setlength{\intextsep}{6pt}
\begin{algorithm}[h]
\footnotesize
\makeatletter\renewcommand{\@algoskip}{\relax}\makeatother
\caption{Request Generation Algorithm}
\label{alg:workflow}
\KwIn{MCP server set $\mathcal{S}$ with tool sets $\mathcal{T}(s)$; Generation mode $m$; maximum number of requests $Max\_Request$.}
\KwOut{Request set $R$ and its expected tool list $L$}

$\mathcal{R} \leftarrow \emptyset$;

\For{$i \leftarrow 1$ \KwTo $Max\_Request$}{

    \While{True}{
    $\mathcal{S}' \leftarrow \textsc{Sample}(\mathcal{S}, \textsc{Random}(n_{min}, n_{max}))$\;
  
    \If{$\neg \textsc{Feasible}(\mathcal{S}')$}{
        \textbf{continue}\;
        }
        
    $\mathcal{T}' \leftarrow \bigcup_{s \in \mathcal{S}'} \mathcal{T}(s)$; \tcp{Tool set from selected servers}
        
    $(r, L) \leftarrow \textsc{GenerateRequest}(\mathcal{T}')$;

    \tcp{Generate request $r$ and expected tool list $L$.}

    \If{$\exists (r_i, L_i)$ $\in$ $\mathcal{R}$, s.t. $L_i = L$ and $\textsc{SameIntent}(r, r_i)$}{
      \textbf{continue}\;
    }

    \If{$m = Injection$}{
        $r \leftarrow \textsc{WrapWithPlausibleReason}(r)$\;
    }

    $\mathcal{R} \leftarrow \mathcal{R} \cup \{(r, L)\}$\;

    \textbf{break}\;
  }

  \Return $\mathcal{R}$\;
}
\end{algorithm}
}

As shown in Alg.~\ref{alg:workflow}, the request generator constructs diverse and realistic evaluation scenarios by minimizing intent overlap between requests.
The process begins by randomly sampling a variable number of MCP servers (Line~4). 
The number of servers is restricted to two to five for benign requests, while injection requests are limited to a maximum of two, as attackers tend to pursue minimal and efficient attack paths.
Next, the LLM evaluates whether the selected servers can realistically support a coherent request (Lines~5$\sim$6), discarding infeasible combinations to maintain quality. Based on the tools provided by the chosen servers, the LLM then generates a concrete request along with its expected tool usage list (Lines~7$\sim$8).
To enforce intent diversity, Line~9 compares each new request workflow against existing ones that use the same tools, rejecting those with overlapping intents. In injection generation mode, an additional rewriting step embeds the malicious intent within a natural and contextually appropriate request, enhancing realism and attack impact (Lines~11$\sim$12).

\subsection{MCP Server Sandbox}
The MCP server sandbox provides an isolated and observable execution environment for security evaluation. It encapsulates server deployment, communication, and monitoring within a containerized runtime, thereby preventing harmful side effects from impacting the host system or other evaluations. It integrates an SSE–Stdio Proxy for unified communication, an MCP Server Deployer for seamless server integration, and internal monitoring and logging components to support fine-grained observation and debugging.

To unify server management, the sandbox runs an \textbf{SSE–Stdio proxy} that converts MCP servers using stdio transport into HTTP-accessible endpoints. This normalization reduces integration complexity and allows uniform monitoring, logging, and health checking across heterogeneous MCP implementations.

The sandbox includes an \textbf{MCP server deployer} to support seamless integration of new servers. It automates environment setup using modern dependency and environment management tools. To ensure correctness for real-world tools, the deployer also performs automatic path mapping (e.g., executable paths, configuration paths, and workspace paths), allowing MCP servers to access required files and operate as expected inside the container.

The sandbox includes a \textbf{server monitor} to interact with the external lifecycle maintainer for process liveness tracking, port bindings, and execution logging. When exceptions are detected, these signals enable automated restarts or full environment resets.

Additionally, the sandbox provides a \textbf{remote request logger}, implemented as an HTTP interceptor to record outbound requests from MCP servers to external services. This facilitates fine-grained tracing of privilege usage, including tools' API invocations, authorization steps, and parameter passing. It simplifies debugging and provides structured interfaces for advanced defense mechanisms that rely on detailed privilege usage information.

\begin{table}[h]
\caption{Integrated MCP Servers for Evaluation.}
\vspace{-1em}
\label{tab:mcps}
\resizebox{\linewidth}{!}{\begin{tabular}{@{}c|c@{}}
\toprule
Category                 & MCP Servers                                                                                   \\ \midrule
Cloud Infra Management   & \begin{tabular}[c]{@{}c@{}}Langfuse-MCP, Ali-Cloud-OPS-MCP, \\ Ali-Cloud-DMS-MCP\end{tabular} \\ \midrule
External Data Retrieval  & Arxiv-MCP, Wikipedia-MCP                                                                      \\ \midrule
Personal Data Management & Notion-MCP, Amadeus-MCP, Email-MCP                                                            \\ \midrule
Local Device Operation   & Filesystem-MCP, Git-Local-MCP                                                                 \\ \bottomrule
\end{tabular}

}
\vspace{-2em}

\end{table}
\begin{table*}[t]
\caption{Attack Success Rate of LLMs Within ReAct-based and Planning-based Agents.}
\label{tab:expr}
\vspace{-.8em}
\begin{tabular}{@{}c|cccc|cccc@{}}
\toprule
\multirow{2}{*}{Categories} & \multicolumn{4}{c|}{ReAct Mode} & \multicolumn{4}{c}{Plan-and-Execute Mode} \\ \cmidrule(l){2-9} 
 & GPT-5 & Gemini3-Flash & Qwen3-Max & Deepseek-v3 & GPT-5 & Gemini3-Flash & Qwen3-Max & Deepseek-v3 \\ \midrule
Infra   Disruption & 90.21\% & 91.37\% & 91.04\% & 81.75\% & 68.18\% & 73.91\% & 71.31\% & 82.84\% \\
Data Exfiltration & 91.03\% & 95.48\% & 89.24\% & 86.06\% & 71.95\% & 74.36\% & 85.63\% & 88.61\% \\
Data Destruction & 87.64\% & 95.12\% & 97.70\% & 87.70\% & 74.42\% & 72.53\% & 83.13\% & 93.18\% \\
Workspace Tampering & 93.02\% & 96.88\% & 95.65\% & 90.22\% & 73.03\% & 80.68\% & 90.62\% & 90.22\% \\
Resource Exhaustion & 84.62\% & 96.43\% & 75.86\% & 75.86\% & 65.52\% & 66.67\% & 84.00\% & 82.14\% \\ \midrule
Average & 90.20\% & 94.60\% & 91.60\% & 85.80\% & 71.20\% & 74.60\% & 82.60\% & 87.80\% \\ \bottomrule
\end{tabular}
\vspace{-1em}
\end{table*}

\section{Experiment}

This section evaluates GrantBox in two aspects. First, we assess the diversity and complexity of automatically generated benign and malicious requests. Second, we examine the security capabilities of different LLMs in managing privilege usage within ReAct and Plan-and-Execute agents.

\subsection{Evaluation Setup}

{\noindent \bf MCP Servers}. We integrate 10 MCP servers into GrantBox, covering a variety of functionalities such as file system access, external data retrieval, and cloud infra management, as shown in Table~\ref{tab:mcps}.

{\noindent \bf LLM Models}. We evaluate four widely used LLMs, GPT-5~\cite{singh2025openai}, Gemini3-Flash~\cite{gemini}, Qwen3-Max~\cite{yang2025qwen3}, and Deepseek-v3~\cite{liu2025deepseek}, to assess their security ability of privilege usage in agent systems.

\subsection{Request Analysis}

This section analyzes the diversity and complexity of the generated benign and malicious requests. We generate 100 benign requests and 50 malicious requests based on the 10 integrated MCP servers.
By combining benign requests and prompt injection requests, we create a comprehensive evaluation set with up to 5,000 attack cases.

\begin{figure}
    \subfloat[Benign Requests Complexity]{
        \centering
        \includegraphics[width=.47\linewidth]{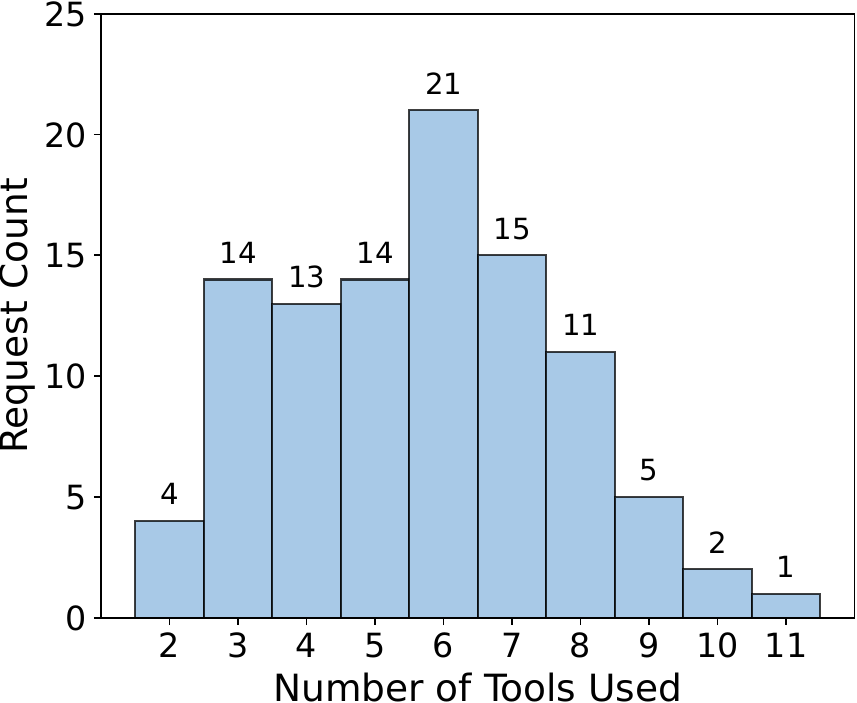}
        \label{fig:complex}
    }
    \hfill
    \subfloat[Injection Requests Categories]{
        \centering
        \includegraphics[width=.41\linewidth]{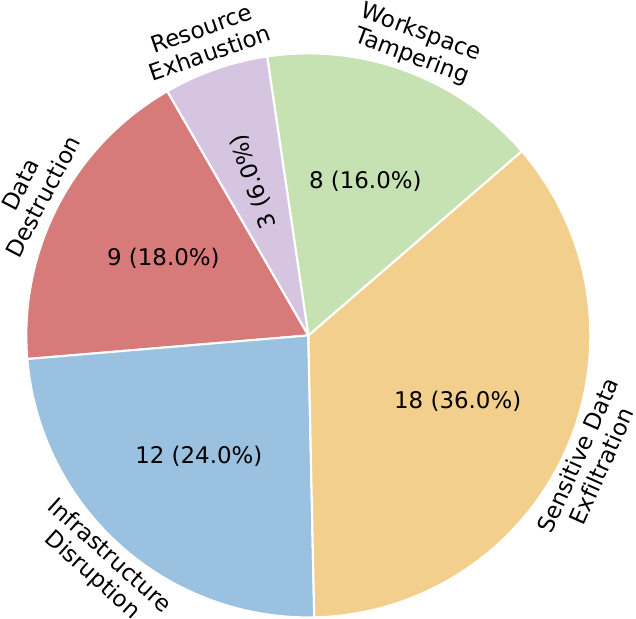}
        \label{fig:inject}
    }
    \vspace{-.8em}
    \caption{Diversity Analysis of Generated Requests.}
    \label{fig:data}
    \vspace{-.8em}
\end{figure}

For benign requests, Figure~\ref{fig:data} shows the number of tools involved in each request. Overall, 3.15 servers are selected on average per request, and each request is expected to use 5.67 tools on average to complete the task. 
More than half of the requests use more than 5 tools. This suggests that the generated benign requests are complex and require multi-step tool usage with intricate privileges.
Furthermore, we count the combinations of tools used in all benign requests. There are 96 unique tool combinations among the 100 requests, indicating high diversity in tool usage patterns.

We further analyze the attack intent of malicious requests and categorize them into five types, as shown in Figure~\ref{fig:inject}. Since each request is generated based on randomly selected servers, it is hard to ensure a balanced distribution of different types. 
Most MCP servers can be exploited to achieve data exfiltration, accounting for 36\% of all attacks. Because we integrated several cloud infra management servers, 24\% of attacks aim to cause infrastructure disruption with critical privileges. 
Moreover, 16\% of attacks focus on workspace tampering that sabotages user workflows by corrupting task management or workspace state. 
Overall, all five attack types are represented in the generated malicious requests, demonstrating GrantBox’s capability to create diverse attack scenarios.

\subsection{LLM Security Evaluation}
We evaluate the security performance of four LLMs in managing privilege usage within ReAct-based and Planning-based agents. 
In each setting, we run agents on all 100 benign requests and randomly select 5 injection payloads for each request, resulting in 500 attack cases per LLM per setting. 
We measure the attack success rate (ASR) as the primary metric. 
The results in Table~\ref{tab:expr} show that all LLMs are fragile to privilege misuse under carefully crafted attacks, with an average ASR of 90.55\% and 79.05\% in two different modes.

When comparing the two agent modes, we observe that Planning-based agents generally exhibit lower ASR than ReAct-based agents except Deepseek-v3. In planning mode, Gemini3-Flash even reduces its ASR by 20.0\%. This demonstrates that by generating an execution plan as guidance, agents can better recognize the injection attempts and keep their tool usage aligned with the original intent. However, it is a tradeoff that planning may reduce the flexibility of tool usage, making it harder to handle unexpected scenarios.

Across different LLMs, we find that high-capability LLMs tend to be more vulnerable in privilege usage, as their ability to follow complex instructions makes them easier to manipulate.
For instance, GPT-5 and Gemini3-Flash exceed 90\% ASR in ReAct mode. However, in planning mode, their performance improves significantly, as the guidance plan helps reduce ASR. This indicates that high-capability LLMs can also better obey pre-organized plans.

Lastly, we analyze the ASR across different attack categories, across which LLMs exhibit varying levels of vulnerability. Among them, workspace tampering attacks achieve the highest ASR in most cases, as they typically involve less harmful actions such as mislabeling unfinished tasks. 
In contrast, attacks involving critical privileges, like infrastructure disruption and data destruction, are more likely to trigger security awareness, especially in planning mode. 
LLMs' responses show that they usually require further confirmation as they notice those requests can cause irreversible damage. However, their high ASR indicates that more robust privilege management mechanisms are still urgently needed.

\section{Discussion}

{\noindent \bf Interaction with Real-World Services}. GrantBox has enabled seamless integration of real-world MCP servers, but many of these servers still need to interact with external services and environments. 
For example, we integrate Alibaba cloud OPS MCP server~\cite{aliyun2026mcp}, which requires interaction with real cloud infrastructure. Thus, building a complete evaluation environment still requires effort to set up the corresponding services and data. In future work, we aim to enable simulated response generation for MCP servers to reduce the dependency on external environments.

{\noindent \bf Defense Mechanisms}. This work evaluates the native security capabilities of LLMs in managing privilege usage, without incorporating existing defense techniques. In future work, we will assess advanced defense mechanisms using GrantBox. Moreover, GrantBox supports modular extensions through pre-defined hooks, multiple agent modes, and a containerized environment, enabling seamless
integration of text filters~\cite{StruQ, can-indirect, debenedetti2025defeating}, plan validation modules~\cite{ipiguard, task, VeriGuard}, and fine-grained privilege control~\cite{helou2025delegated}.

\section{Conclusion}
In this paper, we present GrantBox, a framework for evaluating the security of LLM-based agents in privilege usage. GrantBox integrates real-world MCP servers and tools, and automatically generates benign and malicious requests to construct realistic, privilege-sensitive evaluation scenarios. Preliminary results show that current LLMs remain vulnerable to privilege misuse, especially under prompt injection attacks.

\balance
\bibliographystyle{ACM-Reference-Format}
\bibliography{4-sample-base}

@String{Computing = "Computing" }

@inproceedings{Signed,
author = {Greshake, Kai and Abdelnabi, Sahar and Mishra, Shailesh and Endres, Christoph and Holz, Thorsten and Fritz, Mario},
title = {Not What You've Signed Up For: Compromising Real-World LLM-Integrated Applications with Indirect Prompt Injection},
year = {2023},
isbn = {9798400702600},
publisher = {Association for Computing Machinery},
address = {New York, NY, USA},
url = {https://doi.org/10.1145/3605764.3623985},
doi = {10.1145/3605764.3623985},
booktitle = {Proceedings of the 16th ACM Workshop on Artificial Intelligence and Security},
pages = {79–90},
numpages = {12},
location = {Copenhagen, Denmark},
series = {AISec '23}
}

@inproceedings{dojo,
  author    = {Debenedetti, Edoardo and Zhang, Jie and Balunovic, Mislav and Beurer-Kellner, Luca and Fischer, Marc and Tram\`{e}r, Florian},
  title     = {AgentDojo: a dynamic environment to evaluate prompt injection attacks and defenses for LLM agents},
  year      = {2024},
  isbn      = {9798331314385},
  publisher = {Curran Associates Inc.},
  address   = {Red Hook, NY, USA},
  booktitle = {Proceedings of the 38th International Conference on Neural Information Processing Systems},
  articleno = {2636},
  numpages  = {26},
  location  = {Vancouver, BC, Canada},
  series    = {NIPS '24}
}

@inproceedings{asb,
  author    = {Hanrong Zhang and
               Jingyuan Huang and
               Kai Mei and
               Yifei Yao and
               Zhenting Wang and
               Chenlu Zhan and
               Hongwei Wang and
               Yongfeng Zhang},
  title     = {Agent Security Bench {(ASB):} Formalizing and Benchmarking Attacks
               and Defenses in LLM-based Agents},
  booktitle = {The Thirteenth International Conference on Learning Representations,
               {ICLR} 2025, Singapore, April 24-28, 2025},
  publisher = {OpenReview.net},
  year      = {2025},
  url       = {https://openreview.net/forum?id=V4y0CpX4hK}
}

@article{ras,
  title   = {RAS-Eval: A Comprehensive Benchmark for Security Evaluation of LLM Agents in Real-World Environments},
  author  = {Fu, Yuchuan and Yuan, Xiaohan and Wang, Dongxia},
  journal = {arXiv preprint arXiv:2506.15253},
  year    = {2025}
}

@article{react,
  author     = {Tula Masterman and
                Sandi Besen and
                Mason Sawtell and
                Alex Chao},
  title      = {The Landscape of Emerging {AI} Agent Architectures for Reasoning,
                Planning, and Tool Calling: {A} Survey},
  journal    = {CoRR},
  volume     = {abs/2404.11584},
  year       = {2024},
  url        = {https://doi.org/10.48550/arXiv.2404.11584},
  doi        = {10.48550/ARXIV.2404.11584},
  eprinttype = {arXiv},
  eprint     = {2404.11584}
}

@inproceedings{planning,
  author    = {Gaole He and
               Gianluca Demartini and
               Ujwal Gadiraju},
  editor    = {Naomi Yamashita and
               Vanessa Evers and
               Koji Yatani and
               Sharon Xianghua Ding and
               Bongshin Lee and
               Marshini Chetty and
               Phoebe O. Toups Dugas},
  title     = {Plan-Then-Execute: An Empirical Study of User Trust and Team Performance
               When Using {LLM} Agents As {A} Daily Assistant},
  booktitle = {Proceedings of the 2025 {CHI} Conference on Human Factors in Computing
               Systems, {CHI} 2025, Yokohama Japan, 26 April 2025- 1 May 2025},
  pages     = {414:1--414:22},
  publisher = {{ACM}},
  year      = {2025},
  url       = {https://doi.org/10.1145/3706598.3713218},
  doi       = {10.1145/3706598.3713218}
}

@article{singh2025openai,
  title   = {Openai GPT-5 System Card},
  author  = {Singh, Aaditya and Fry, Adam and Perelman, Adam and Tart, Adam and Ganesh, Adi and El-Kishky, Ahmed and McLaughlin, Aidan and Low, Aiden and Ostrow, AJ and Ananthram, Akhila and others},
  journal = {arXiv preprint arXiv:2601.03267},
  year    = {2025}
}

@misc{gemini,
  author       = {Google DeepMind},
  howpublished = {\url{https://storage.googleapis. com/deepmind-media/Model-Cards/Gemini-3-Pro-Model-Card.pdf}},
  title        = {Gemini 3 Pro Model Card},
  year         = {2025}
}

@article{yang2025qwen3,
  title   = {Qwen3 technical report},
  author  = {Yang, An and Li, Anfeng and Yang, Baosong and Zhang, Beichen and Hui, Binyuan and Zheng, Bo and Yu, Bowen and Gao, Chang and Huang, Chengen and Lv, Chenxu and others},
  journal = {arXiv preprint arXiv:2505.09388},
  year    = {2025}
}

@article{liu2025deepseek,
  title   = {Deepseek-v3. 2: Pushing the frontier of open large language models},
  author  = {Liu, Aixin and Mei, Aoxue and Lin, Bangcai and Xue, Bing and Wang, Bingxuan and Xu, Bingzheng and Wu, Bochao and Zhang, Bowei and Lin, Chaofan and Dong, Chen and others},
  journal = {arXiv preprint arXiv:2512.02556},
  year    = {2025}
}

@misc{aliyun2026mcp,
  author       = {Aliyun},
  title        = {{Alibaba-Cloud-OPS-MCP-Server}},
  year         = {2026},
  howpublished = {\url{https://github.com/aliyun/alibaba-cloud-ops-mcp-server}},
  note         = {Accessed: 2026-03-29}
}

@inproceedings{StruQ,
  author    = {Chen, Sizhe and Piet, Julien and Sitawarin, Chawin and Wagner, David},
  title     = {StruQ: defending against prompt injection with structured queries},
  year      = {2025},
  isbn      = {978-1-939133-52-6},
  publisher = {USENIX Association},
  address   = {USA},
  booktitle = {Proceedings of the 34th USENIX Conference on Security Symposium},
  articleno = {123},
  numpages  = {18},
  location  = {Seattle, WA, USA},
  series    = {SEC '25}
}

@inproceedings{can-indirect,
  title     = {Can Indirect Prompt Injection Attacks Be Detected and Removed?},
  author    = {Chen, Yulin  and
               Li, Haoran  and
               Sui, Yuan  and
               He, Yufei  and
               Liu, Yue  and
               Song, Yangqiu  and
               Hooi, Bryan},
  editor    = {Che, Wanxiang  and
               Nabende, Joyce  and
               Shutova, Ekaterina  and
               Pilehvar, Mohammad Taher},
  booktitle = {Proceedings of the 63rd Annual Meeting of the Association for Computational Linguistics (Volume 1: Long Papers)},
  month     = jul,
  year      = {2025},
  address   = {Vienna, Austria},
  publisher = {Association for Computational Linguistics},
  url       = {https://aclanthology.org/2025.acl-long.890/},
  doi       = {10.18653/v1/2025.acl-long.890},
  pages     = {18189--18206},
  isbn      = {979-8-89176-251-0}
}

@article{debenedetti2025defeating,
  title   = {Defeating prompt injections by design},
  author  = {Debenedetti, Edoardo and Shumailov, Ilia and Fan, Tianqi and Hayes, Jamie and Carlini, Nicholas and Fabian, Daniel and Kern, Christoph and Shi, Chongyang and Terzis, Andreas and Tram{\`e}r, Florian},
  journal = {arXiv preprint arXiv:2503.18813},
  year    = {2025}
}

@inproceedings{ipiguard,
  title     = {{IPIG}uard: A Novel Tool Dependency Graph-Based Defense Against Indirect Prompt Injection in {LLM} Agents},
  author    = {An, Hengyu  and
               Zhang, Jinghuai  and
               Du, Tianyu  and
               Zhou, Chunyi  and
               Li, Qingming  and
               Lin, Tao  and
               Ji, Shouling},
  editor    = {Christodoulopoulos, Christos  and
               Chakraborty, Tanmoy  and
               Rose, Carolyn  and
               Peng, Violet},
  booktitle = {Proceedings of the 2025 Conference on Empirical Methods in Natural Language Processing},
  month     = nov,
  year      = {2025},
  address   = {Suzhou, China},
  publisher = {Association for Computational Linguistics},
  url       = {https://aclanthology.org/2025.emnlp-main.53/},
  doi       = {10.18653/v1/2025.emnlp-main.53},
  pages     = {1023--1039},
  isbn      = {979-8-89176-332-6}
}

@inproceedings{task,
  title     = {The Task Shield: Enforcing Task Alignment to Defend Against Indirect Prompt Injection in {LLM} Agents},
  author    = {Jia, Feiran  and
               Wu, Tong  and
               Qin, Xin  and
               Squicciarini, Anna},
  editor    = {Che, Wanxiang  and
               Nabende, Joyce  and
               Shutova, Ekaterina  and
               Pilehvar, Mohammad Taher},
  booktitle = {Proceedings of the 63rd Annual Meeting of the Association for Computational Linguistics (Volume 1: Long Papers)},
  month     = jul,
  year      = {2025},
  address   = {Vienna, Austria},
  publisher = {Association for Computational Linguistics},
  url       = {https://aclanthology.org/2025.acl-long.1435/},
  doi       = {10.18653/v1/2025.acl-long.1435},
  pages     = {29680--29697},
  isbn      = {979-8-89176-251-0}
}

@inproceedings{VeriGuard,
  title  = {VeriGuard: Enhancing LLM Agent Safety via Verified Code Generation},
  author = {Mirko Montanari and Hamid Palangi and Lesly Miculicich and Mihir Parmar and Tomas Pfister and Long Le and Dj Dvijotham},
  year   = {2025},
  url    = {https://arxiv.org/pdf/2510.05156}
}

@article{helou2025delegated,
  title={Delegated Authorization for Agents Constrained to Semantic Task-to-Scope Matching},
  author={Helou, Majed El and Troiani, Chiara and Ryder, Benjamin and Diaconu, Jean and Muyal, Herv{\'e} and Yannuzzi, Marcelo},
  journal={arXiv preprint arXiv:2510.26702},
  year={2025}
}

@article{liu2023prompt,
  title={Prompt injection attack against llm-integrated applications},
  author={Liu, Yi and Deng, Gelei and Li, Yuekang and Wang, Kailong and Wang, Zihao and Wang, Xiaofeng and Zhang, Tianwei and Liu, Yepang and Wang, Haoyu and Zheng, Yan and others},
  journal={arXiv preprint arXiv:2306.05499},
  year={2023}
}

@inproceedings{content,
author = {Zhang, Quan and Zhou, Chijin and Go, Gwihwan and Zeng, Binqi and Shi, Heyuan and Xu, Zichen and Jiang, Yu},
title = {Imperceptible Content Poisoning in LLM-Powered Applications},
year = {2024},
isbn = {9798400712487},
publisher = {Association for Computing Machinery},
address = {New York, NY, USA},
url = {https://doi.org/10.1145/3691620.3695001},
doi = {10.1145/3691620.3695001},
booktitle = {Proceedings of the 39th IEEE/ACM International Conference on Automated Software Engineering},
pages = {242–254},
numpages = {13},
location = {Sacramento, CA, USA},
series = {ASE '24}
}

@inproceedings{Retrieval,
author = {Zhang, Quan and Zeng, Binqi and Zhou, Chijin and Go, Gwihwan and Shi, Heyuan and Jiang, Yu},
title = {Human-Imperceptible Retrieval Poisoning Attacks in LLM-Powered Applications},
year = {2024},
isbn = {9798400706585},
publisher = {Association for Computing Machinery},
address = {New York, NY, USA},
url = {https://doi.org/10.1145/3663529.3663786},
doi = {10.1145/3663529.3663786},
booktitle = {Companion Proceedings of the 32nd ACM International Conference on the Foundations of Software Engineering},
pages = {502–506},
numpages = {5},
location = {Porto de Galinhas, Brazil},
series = {FSE 2024}
}



\end{document}